# Performance Optimization of MapReduce-based Apriori Algorithm on Hadoop Cluster

Sudhakar Singh [a, *], Rakhi Garg [b], P K Mishra [c]

[a] Department of Computer Science, Institute of Science, Banaras Hindu University, Varanasi 221005, India
[b] Mahila Maha Vidyalaya, Banaras Hindu University, Varanasi 221005, India
[c] Department of Computer Science & DST-Centre for Interdisciplinary Mathematical Sciences, Banaras Hindu University, Varanasi 221005, India

**ABSTRACT**

Many techniques have been proposed to implement the Apriori algorithm on MapReduce framework but only a few have focused on performance improvement. FPC (Fixed Passes Combined-counting) and DPC (Dynamic Passes Combined-counting) algorithms combine multiple passes of Apriori in a single MapReduce phase to reduce the execution time. In this paper, we propose improved MapReduce based Apriori algorithms VFPC (Variable Size based Fixed Passes Combined-counting) and ETDPC (Elapsed Time based Dynamic Passes Combined-counting) over FPC and DPC. Further, we optimize the multi-pass phases of these algorithms by skipping pruning step in some passes, and propose Optimized-VFPC and Optimized-ETDPC algorithms. Quantitative analysis reveals that counting cost of additional un-pruned candidates produced due to skipped-pruning is less significant than reduction in computation cost due to the same. Experimental results show that VFPC and ETDPC are more robust and flexible than FPC and DPC whereas their optimized versions are more efficient in terms of execution time.

## 1. INTRODUCTION

The Apriori algorithm proposed by R. Agrawal and R. Srikant [1] is one of the most popular and widely used data mining algorithms that mines frequent itemsets using candidate generation. Apriori is the basic algorithm of Association Rule Mining (ARM) and its genesis boosted the research in data mining. Apriori is one among the top 10 data mining algorithms identified by the IEEE International Conference on Data Mining (ICDM) in 2006 on the basis of most influential data mining algorithms in the research community [2].

---

\* Corresponding Author
*E-mail addresses:* sudhakarcsbhu@gmail.com (Sudhakar Singh), rgarg@bhu.ac.in (Rakhi Garg), mishra@bhu.ac.in (P K Mishra)

Data does not remain only large in volume but also exhibits other characteristic like velocity i.e. data in motion and variety i.e. data in many form. The recent trend is of Big Data [3] which is majorly defined by high volume, high velocity and high variety. The traditional data mining techniques and tools are efficient in analyzing/mining data but not scalable and efficient in managing big data. Big data architectures and technologies are adopted to analyze such data. Hadoop is a large-scale distributed batch processing infrastructure for parallel processing of big data on large cluster of commodity computers [4]. MapReduce is a parallel programming model of Hadoop designed for parallel processing of large volumes of data. Therefore, it is required to redesign the data mining algorithms on MapReduce framework in order to mine big data sets [5].

In MapReduce programming model, an application is called a MapReduce Job which consists of Mapper and Reducer and input datasets are stored in Hadoop Distributed File System (HDFS). The entire map and reduce tasks are executed on different machines in parallel fashion but the final result is obtained only after the completion of all the reduce tasks. If algorithm is recursive, then we have to execute multiple MapReduce jobs to get the final result [6]. To redesign the Apriori algorithm on MapReduce framework we have to define two independent methods map and reduce and to convert input datasets in the form of (key, value) pairs. Apriori algorithm is an iterative process and its two main components are candidate itemsets generation and frequent itemsets generation. In each job, database is scanned, Mapper generates local candidates, and Reducer sums up the local count and results frequent itemsets.

The straight forward implementation of Apriori on MapReduce framework adds the overheads of scheduling and waiting time. Due to purely iterative nature of Apriori, a new job is invoked each time for the new iteration and also the next job cannot be started until previous job has finished. M-Y. Lin *et al.* [7] proposed three version of Apriori on MapReduce, named Single Pass Counting (SPC), Fixed Passes Combined-counting (FPC) and Dynamic Passes Combined-counting (DPC). Algorithms FPC and DPC considerably improve the performance, and are the most efficient implementation of Apriori on MapReduce framework with the minimum number of iterations. SPC is a straight forward implementation of Apriori on MapReduce whereas FPC and DPC combine generation and counting of candidates of multiple consecutive phases/jobs of SPC in a single phase/job. FPC combines fixed number of consecutive phases of SPC (generally 3 phases) into a single MapReduce phase, reducing the number of scheduling invocations. DPC dynamically merges the candidates of several consecutive phases to balance the workloads between phases [7]. Now suppose we combined candidate generation of 3 consecutive passes *k, k+1, k+2* in a single multi-pass phase, then candidate k-itemsets are generated from frequent (k-1)-itemsets, candidate (k+1)-itemsets from candidate k-itemsets and candidate (k+2)-itemsets from candidate (k+1)-itemsets. When a candidate itemsets of next level are generated from candidate itemsets itself instead of frequent itemsets then it produces some false-positive candidates [7]. FPC might perform poorly due to the overloaded false-positive candidates and it happens since FPC combines the same fixed number of passes in all phases. DPC overcomes this problem by dynamic combination of consecutive passes. The drawback of DPC is the strategy it uses to determine dynamic number of passes to combine. DPC is directly dependent on the execution time of preceding phase to decide combination of passes. Execution time cannot be the absolute parameter since it may vary in clusters of different size and capacity as well as on new datasets.

## 1.1. Contributions

In this paper, we have proposed four algorithms, named VFPC (Variable Size based Fixed Passes Combined-counting), ETDPC (Elapsed Time based Dynamic Passes Combined-counting), Optimized-VFPC and Optimized-ETDPC. VFPC and ETDPC are improvement over existing FPC and DPC algorithms [7]. FPC and DPC are well known efficient implementation of Apriori on MapReduce framework. VFPC combines candidates of variable number of consecutive passes in different phases. It combines small number of passes (generally 2) in earlier phases and increases the number of passes in latter phases (e.g. 4, 6…). When to increase the number of passes to be combined is determined by the number of candidates in two preceding consecutive phases. ETDPC combines candidates of dynamic number of consecutive passes in different phases. The number of passes to be combined depends on the elapsed time of two preceding consecutive phases. Further, in combined passes when candidates are generated from candidates itself, if the pruning step is skipped then it produces some un-pruned candidates without loss of integrity of frequent itemsets produced at the end. So, we have proposed optimized multi-pass phase that uses pruning step in first pass and skips pruning in all the latter passes in that phase. Optimized-VFPC and Optimized-ETDPC are similar to VFPC and ETDPC but with optimized phases. We have executed our proposed algorithms on both real life and synthetic datasets and found that VFPC and ETDPC are more robust and flexible to the new datasets and Hadoop clusters with different computing capacity in comparison to FPC and DPC. In terms of performance, a significant reduction in execution time of Optimized-VFPC and Optimized-ETDPC has been found in comparison to VFPC and ETDPC. Moreover, the proposed algorithms are most effective for long frequent itemset mining. Longer frequent itemsets are generated either in dense datasets or at the lower minimum support.

The rest of this paper is organized as follows. Section 2 describes the fundamental concepts of Apriori algorithm and Hadoop system. Section 3 briefly reviews the related works. Our proposed algorithms are described and analyzed in section 4. Experimental results are discussed in section 5. Finally section 6 concludes the paper.

## 2. BASIC CONCEPTS

## 2.1. Apriori Algorithm

Apriori is an iterative algorithm which alternates between the two important tasks, the first one is generation of candidates from frequent itemsets of previous iteration and the second one is scanning of database for support counting of candidates against each transaction. In $k^{th}$ iteration ($k \geq 2$), candidate k-itemsets $C_k$ is generated from frequent (k-1)-itemsets $L_{k-1}$ and then k-itemset subsets of each transaction is checked against candidates in $C_k$ for support counting. Candidate itemsets $C_k$ is obtained by conditionally joining $L_{k-1}$ with itself and then pruning those itemsets that does not satisfy Apriori property. According to this property all the itemsets of $C_k$ can be removed from $C_k$ if anyone of their (k-1)-subsets does not present in $L_{k-1}$ [1].

## 2.2. Hadoop Distributed File System and MapReduce

Hadoop integrates the computational power (MapReduce) with distributed storage (Hadoop Distributed File System) to reduce the communication cost by providing local access to data and local computation on data. Hadoop Distributed File System (HDFS) is designed based on GFS (Google File System) to be deployed on low-cost hardware. It breaks files in fixed size blocks (default block size is 64 MB) and replicates (default replication factor is 3) the blocks across multiple machines in cluster to provide high availability and fault tolerance [8].

MapReduce [9] is an efficient, scalable and simplified programming model for large scale distributed data processing on a large cluster of commodity computers. It process large volumes of data in parallel by breaking the job into independent tasks across a large number of machines. MapReduce program generally consists of Mapper, Combiner and Reducer tasks, which runs on all machines in a Hadoop cluster. The input and output of these functions must be in the form of *(key, value)* pairs [8]. Followings are the specific details of MapReduce paradigm followed in our implementations.

An InputFormat class selects input file residing in HDFS, defines the InputSplit that splits the file and provides RecordReaders that reads each split for individual mapper. A single map task is a unit of work in a MapReduce program which corresponds to a single InputSplit. RecordReader reads the splitted data and converts it into *(key, value)* pairs for the mapper. The Mapper takes the input $(k_1, v_1)$ pairs, performs the user defined computation and produces a list of intermediate $(k_2, v_2)$ pairs. These pairs are partitioned per reducer by the Partitioner class. The Reducer takes $(k_2, list(v_2))$ pairs as input, make sum of the values in *list $(v_2)$* and produce new pairs $(k_2, v_3)$. Each reducer writes its output in a separate file in HDFS which is directed by OutputFormat class. An optional Combiner class is used to reduce the communication cost of transferring intermediate output of mappers to reducers. It is known as mini reducer since it works locally on all the *(key, value)* pair emitted by the map tasks on a given node only [8].

## 3. RELATED WORKS

To enhance the performance and scalability of association rule mining algorithms, many parallel and distributed algorithms have been developed for homogeneous computing environment as well as for heterogeneous environment like grid computing [10]. No doubt these parallel and distributed algorithms improve the mining performance but involve overheads of managing parallel and distributed systems. These overhead are computation partitioning, data partitioning, synchronization, communication, scheduling, work load balancing and managing nodes failure in cluster or grid [7, 11]. All these problems can be overcome by the MapReduce framework originally introduced by Google [9].

Several frequent itemset mining algorithms have been proposed on MapReduce framework since the foundation of Hadoop. Among these proposed algorithms, majors are Apriori based [6, 7] [12-16] while some are FP-Growth based [17] and very few are based on Eclat [11]. Most of the Apriori based algorithms are simply straight forward implementation of Apriori on MapReduce framework [12]. The algorithms FPC and DPC [7] significantly improve the performance of MapReduce based Apriori by proposing an innovative idea of combining multiple consecutive iterations of Apriori in a single MapReduce Job. F. Kovacs and J. Illes [6] used the functionality of the Combiner inside the Mapper and candidate generation inside the Reducer instead of the Mapper. They also proposed a technique to count 1 and 2–itemsets in one step using triangular matrix data structure. L. Li and M. Zhang [13] proposed a method of dataset distribution suitable for Hadoop cluster consisting of heterogeneous nodes, and used it in Apriori implementation. A parallel randomized algorithm, PARMA is proposed by Matteo Riondato *et al.* [14] on MapReduce for discovering approximate frequent itemsets. It mines a small random sample of the datasets, and is independent from the dataset size. T. Y. Jen *et al.* [15] have used the vertical data layout for Apriori to reduce some operations and improve the scalability and efficiency. The influence of three data structures hash tree, trie, and hash table trie (trie with hashing technique) on MapReduce based Apriori algorithm have been investigated in [16] and it has been found that hash table trie drastically outperforms trie and hash tree. Y. Xun *et al.* [18] have developed a parallel frequent itemset mining algorithm on MapReduce called FiDoop. It incorporates FIU-tree (frequent items ultrametric tree) in place of FP tree [17]. Its extension called FiDoop-HD has also been designed for processing high dimensional data. The review paper [19] describes in detailed about the MapReduce based Apriori and compares the algorithmic features of various MapReduce based Apriori algorithms.

Although, the work in this paper is centered to MapReduce based Apriori on Hadoop, but there are some other platforms that have been emerged after Hadoop. Apache's Mahout [20] provides the library of various machine learning and data mining algorithms. Machine learning library of Mahout and that of Spark (*spark.mllib*) [21] both provide the implementation of Parallel FP-Growth algorithm only, based on the paper [17]. Apache's Spark [21] processes data in memory and executes job 10 to 100 times faster than MapReduce. Spark uses a lot of memory due to in memory computation, so it needs a dedicated high end physical machine while Hadoop works very well on a commodity machine. Further, Spark also uses the concept of Map and Reduce functions along with its own other functions. R-Apriori [22] is one of the efficient implementations of parallel Apriori algorithm on Spark.

Mining data from real-time transactions requires big data stream computing (BDSC) [23] that process large volume of data at a high speed in real time. A new paradigm of big data stream mobile computing (BDSMC) has been formalized in [24]. It has focus on the real-time processing and energy efficiency for managing computing-communication platforms supporting BDSMC. Another relevant area, fuzzy system based data mining [25] enables to indentify the imprecise relations among the items of database. MapReduce based algorithms have also been designed for mining fuzzy association rules [26].

## 4. PROPOSED ALGORITHMS

In MapReduce based Apriori one need to design a Mapper which generates the candidates with local support count for its assigned split of database, and a Reducer to sum up the local counts and generates frequent itemsets with global support count. A Combiner may be used to minimize the communication cost between Mappers and Reducers. The design of Combiner and the Reducer remains the same in all variations of MapReduce based Apriori since main function of both is to make sum of local count. The design of the Mapper varies depending on the required computations. In frequent 1-itemsets generation the Mapper invokes its map method for each transaction from block of database assigned to the Mapper. Map method outputs *(item, 1)* pairs for each item in the transaction. Mapper for computing 1-itemset is named as OneItemsetMapper while Combiner and Reducer are named as ItemsetCombiner and ItemsetReducer respectively. Algorithm 1 depicts the pseudo codes for Mapper generating 1-itemset, and Combiner and Reducer. The algorithms for ItemsetCombiner and ItemsetReducer are the same except that former does not check support count against minimum support threshold. A very detailed description of straight forward implementation of MapReduce based Apriori is represented in [19].

**Algorithm 1** OneItemsetMapper, ItemsetCombiner and ItemsetReducer

| **OneItemsetMapper, k = 1** | **ItemsetCombiner** | **ItemsetReducer** |
|---|---|---|
| Input: a block $b_i$ of database | key: itemset | key: itemset |
| key: byte offset of the line | value: key's value list | value: key's value list |
| value: a transaction $t_i \in b_i$ | 1: function reduce(key, value) | 1: function reduce(key, value) |
| 1: function map(key, value) | 2:    for each value v of key's value list | 2:    for each value v of key's value list |
| 2:    for each item I $\in t_i$ do | 3:      sum += v; | 3:      sum += v; |
| 3:      write (I, 1); | 4:    end for | 4:    end for |
| 4:    end for | 5:    write(itemset, sum) | 5:    if sum >= min_sup_count |
| 5: end function map | 6: end function reduce | 6:      write(itemset, sum) |
|  |  | 7:    end if |
|  |  | 8: end function reduce |

**Algorithm 2** SPC with SPCItemsetMapper

| **SPCItemsetMapper, k ≥ 2** | **SPC Driver** |
|---|---|
| Input: a block $b_i$ of database and $L_{k-1}$ | Find frequent 1-itemset $L_1$ |
| key: byte offset of the line | 1: Job1 |
| value: a transaction $t_i \in b_i$ | 2:    OneItemsetMapper |
| 1: read $L_{k-1}$ from cache file in a prefix tree trie$L_{k-1}$ | 3:    ItemsetCombiner |
| 2: function map(key, value) | 4:    ItemsetReducer |
| 3:    trie$C_k$ = apriori-gen(trie$L_{k-1}$); | 5: end Job1 |
| 4:    $C_t$ = subset(trie$C_k$, $t_i$); | Find frequent k-itemset $L_k$ |
| 5:    for each candidate c $\in C_t$ do | 6: for (k = 2; $L_{k-1} \neq \phi$; k++) |
| 6:      write (c, 1); | 7:    Job2 |
| 7:    end for | 8:      SPCItemsetMapper |
| 8: end function map | 9:      ItemsetCombiner |
|  | 10:     ItemsetReducer |
|  | 11:   end Job2 |
|  | 12: end for |

We have used two types of MapReduce Job named as Job1 and Job2. Job 1 generates frequent 1-itemsets and Job2 generates k-itemsets (*k ≥ 2*). Job1 is configured of OneItemsetMapper, ItemsetCombiner and ItemsetReducer, which remains the same for all the algorithms. Single Pass Counting (SPC) algorithm [7] is a straight forward implementation of Apriori on MapReduce. We named its Mapper as SPCItemsetMapper that is used in Job2. Job2 of SPC is configured of SPCItemsetMapper, ItemsetCombiner and ItemsetReducer. SPC invokes new instance of Job2 each time for each single pass/iteration of Apriori. Pseudo code of SPCItemsetMapper and SPC is shown in Algorithm 2. SPCItemsetMapper makes use of *apriori-gen()* and *subset()* methods of Apriori [1] for generating candidates and support counting. In our implementations, we have used the Prefix Tree (Trie) data structure [27] in all the algorithms for storing and generating candidates. Prefix tree for frequent k-itemsets and candidate k-itemsets are represented by $trieL_k$ and $trieC_k$ respectively.

The existing FPC and DPC algorithms [7] combine the consecutive MapReduce phases of SPC (i.e. passes of Apriori) in a single phase to reduce the scheduling invocations and waiting time. FPC combines fixed number of passes due to which it suffers with the overloaded false positive candidates in early phases while latter phases compute very less candidates. DPC overcomes this problem and dynamically combines the passes to balance the workload in each phase. As many passes are combined dynamically such that the total number of candidates generated in these passes cannot exceed a candidate threshold value. The candidate threshold $ct$ is defined as $ct = \alpha \times |L_{k-1}|$ where $|L_{k-1}|$ is the number of the longest sized frequent itemsets of previous phase and $\alpha$ is determined heuristically. The value of $\alpha$ may set to higher than 1 (e.g. 1.2 or 2) if execution time of previous phase is less than a threshold value $\beta$ (e.g. $\beta = 60$ sec.) otherwise set to 1. Execution time cannot be an absolute criterion since it is different for the same algorithm on different clusters or for new datasets. We have proposed four algorithms VFPC, ETDPC, Optimized-VFPC and Optimized-ETDPC to overcome the problems with FPC and DPC. VFPC and ETDPC are more robust and flexible than FPC and DPC for new datasets as well as for clusters with different computing capacity. Optimized-VFPC and Optimized-ETDPC are optimized version of VFPC and ETDPC based on skipped-pruning.

### 4.1. VFPC (Variable Size based FPC) and ETDPC (Elapsed Time based DPC)

VFPC and ETDPC are based on the simple strategy of combining lesser number of passes in earlier phases and more number of passes in latter phases. In Apriori algorithm, it has been observed that initially the number of candidate/frequent itemsets are small (e.g. in initial passes, k=1, 2), it increases as the iteration counter increases and after a certain iteration it starts to decrease. In other words, the number of candidate/frequent itemsets in starting and ending iterations are quantitatively smaller than in middle of iterations. Number and width of the itemsets greatly influence the computation time. It has also been observed that execution time of starting and ending iterations are much smaller than that of middle of iterations. So if we could have identified the certain point before that the number of candidates and execution time per iteration are increasing with iteration counter and after that point starts decreasing; then we can combine the consecutive passes accordingly to balance the workload among phases. We have formulated two techniques based on number of candidates and elapsed time to efficiently combine the passes, which are the foundation of VFPC and ETDPC respectively. Algorithms 3 and 4 represent the pseudo codes of VFPC and ETDPC respectively along with their corresponding Mappers.

VFPC combines candidates of *npass* consecutive passes in one phase (Algorithm 3). Initially *npass* is equal to 2 and remains the same until the number of candidates per phase is increasing. VFPC keeps track of the number of candidates in two preceding consecutive phases and compares them. Two variables *numCandsK* and *numCandsKprev* represent the number of candidates in phase currently completed and phase just previous to it respectively. VFPCItemsetMapper combines only two passes until *numCandsK* is greater than or equal to *numCandsKprev*. When *numCandsK* becomes less than *numCandsKprev* then the number of passes to be combined is incremented by 3 in further phase. So, VFPC combines fixed number of passes in each phase but that fixed number is not the same for all phases. It combines only 2 passes in earlier phases (k ≥ 2) to avoid candidates overloading and 5 or more passes in latter phases.

| **Algorithm 3** VFPC with VFPCItemsetMapper | |
|---|---|
| **VFPCItemsetMapper, k ≥ 2** | **VFPC Driver** |
| Input: a block $b_i$ of database and $L_{k-1}$ | Find frequent 1-itemset $L_1$ |
| key: byte offset of the line | 1: Job1 // same as Job1 of SPC |
| value: a transaction $t_i \in b_i$ | Find frequent k-itemset $L_k$ |
| 1: read $L_{k-1}$ from cache file in a prefix tree trie$L_{k-1}$ | 2: k = 2, npass = 2, numCandsKprev = 0; |
| 2: get the value of k and npass from context | 3: do while($L_{k-1} \neq \phi$) |
| 3: function map(key, value) | 4:    Job2 |
| 4:    trieC$_k$ = trieL$_{k-1}$; | 5:    set the value of k and npass to job configuration |
| 5:    candidateCount = 0; | 6:    VFPCItemsetMapper |
| 6:    for (count = 1; count ≤ npass; count++) | 7:    ItemsetCombiner |
| 7:      trieC$_k$ = apriori-gen(trieC$_k$); | 8:    ItemsetReducer |
| 8:      C$_t$ = subset(trieC$_k$, t$_i$); | 9:    end Job2 |
| 9:      for each candidate c ∈ C$_t$ do | 10:    numCandsK = candidateCount |
| 10:        write (c, 1); | 11:    if(numCandsK < numCandsKprev) |
| 11:      end for | 12:      npass += 3; |
| 12:      candidateCount += |trieC$_k$| ; | 13:    else |
| 13:    end for | 14:      npass = 2; |
| 14:    set the value of candidateCount to context | 15:    end if-else |
| 15: end function map | 16:    numCandsKprev = numCandsK, k += npass; |
| | 17: end do-while |

ETDPC dynamically determines the number of consecutive passes to be combined such that the total number of candidates generated in combined passes cannot exceed a candidate threshold value, $ct = \alpha \times |L_{k-1}|$, where $|L_{k-1}|$ is the number of the longest sized frequent itemsets of previous phase and the value of $\alpha$ is determined by elapsed time of two preceding consecutive phases of ETDPC (Algorithm 4). Here we define two time limits $\beta_1$ and $\beta_2$ as 40 seconds and 60 seconds respectively. Two variables *ET* and *ETprev* represent the elapsed time of phase currently completed and phase just previous to it respectively. We decide the value of $\alpha$ on the basis of $\beta_1$ and $\beta_2$ until *ET* is greater than *ETprev*. The value of $\alpha$ is set to 3 if *ET* is less than or equal to $\beta_1$ else set to 2 if *ET* is less than $\beta_2$ and greater than $\beta_1$ otherwise it is set to 1. When *ETprev* becomes greater than or equal to *ET* then value of $\alpha$ is decided on the basis of time difference between *ETprev* and *ET*. The value of $\alpha$ is set to 3 if *ETprev* is greater than or equal to 1.5 times of *ET* otherwise it is set to 2. ETDPCItemsetMapper calculates the value of candidate threshold *ct* and combined the passes accordingly.

| **Algorithm 4** ETDPC with ETDPCItemsetMapper | |
|---|---|
| **ETDPCItemsetMapper, k ≥ 2** | **ETDPC Driver** |
| Input: a block $b_i$ of database and $L_{k-1}$ | Find frequent 1-itemset $L_1$ |
| key: byte offset of the line | 1: Job1 // same as Job1 of SPC |
| value: a transaction $t_i \in b_i$ | Find frequent k-itemset $L_k$ |
| 1: read $L_{k-1}$ from cache file in a prefix tree $trieL_{k-1}$ | 2: k = 2, α = 1, $β_1$ = 40; $β_2$ = 60, npass = 1; |
| 2: get the value of k and α from context | 3: ETprev = elapsed time of Job1; |
| 3: function map(key, value) | 4: do while ($L_{k-1} \ne \phi$) |
| 4:    candidate threshold ct = α * \|$L_{k-1}$\|; | 5:    Job2 |
| 5:    $trieC_k$ = $trieL_{k-1}$; | 6:      set the value of k and α to job configuration |
| 6:    candidateCount = 0; | 7:      ETDPCItemsetMapper |
| 7:    npass = 0; | 8:      ItemsetCombiner |
| 8:    do while(candidateCount ≤ ct) | 9:      ItemsetReducer |
| 9:      $trieC_k$ = apriori-gen($trieC_k$); | 10:    end Job2 |
| 10:      $C_t$ = subset($trieC_k$, $t_i$); | 11:    update the value of npass; |
| 11:      for each candidate c ∈ $C_t$ do | 12:    ET = elapsed time of Job2; |
| 12:        write (c, 1); | 13:    if(ETprev < ET) |
| 13:      end for | 14:      if(ET ≤ $β_1$)  α = 3; |
| 14:      candidateCount += \|$trieC_k$\| ; | 15:      else if(ET < $β_2$)  α = 2; |
| 15:      npass++; | 16:      else  α = 1; |
| 16:    end do-while | 17:      end if-else |
| 17:    set the value of npass to context | 18:    else if(ETprev ≥ ET) |
| 18: end function map | 19:      if(ETprev ≥ 1.5 * ET)  α = 3; |
| | 20:      else  α = 2; |
| | 21:      end if-else |
| | 22:    end if-else |
| | 23:    ETprev = ET, k += npass; |
| | 24: end do-while |

Advantage with VFPC and ETDPC is that they support robustness and flexibility against new datasets and Hadoop clusters with different computing capacity. On some datasets, it is empirically observed that FPC performs very poorly and converges to SPC at lower *min_sup* (Section 5) due to combining the same number of passes in all phases. VFPC is robust for the new datasets and do not converge to SPC. VFPC combines two passes in starting phases that avoids overloaded phase while almost all the latter passes that generates very small number of candidates are combined in one phase. On the other hand, DPC overcomes the problem of overloaded phases in FPC but the solution it provides is based on the execution time of previous phase and a threshold value *β*. Execution time of previous phase may not be uniform on new cluster with lower or higher computing power and so the value of *β* need to be changed every time to fine tune the performance. ETDPC is flexible and do not require such adjustment. It also avoids overloaded phase and combines all the latter passes in minimum number of phases at the end. It combines the passes based on the relative elapsed time of preceding consecutive phases rather than only elapsed time of previous phase.

### 4.2. Optimized-VFPC and Optimized-ETDPC

In optimized version of VFPC and ETDPC, we have optimized the computations by skipping the pruning step in some of the combined passes in the multi-pass phases. Pruning step reduces the number of candidates by applying Apriori property. The respective Mappers of VFPC and ETDPC use *apriori-gen()* method to generate candidate itemsets. The method *apriori-gen()* consists of two tasks join and prune. We have designed a new method *non-*

*apriori-gen()* which skips the pruning step and consists of merely join step. We have employed these two candidate generation methods in the multi-pass phases of Optimized-VFPC and Optimized-ETDPC algorithms. The method *non-apriori-gen()* produces increased number of candidate itemsets since it includes un-pruned candidates. Using *non-apriori-gen()* method saves the cost of pruning but adds the counting of additional un-pruned candidates. In the following sub-section, we have analyzed the effect of skipped-pruning in a multi-pass phase. Algorithm 5 depicts the pseudo code of the Mappers of Optimized-VFPC and Optimized-ETDPC named as Optimized-VFPCItemsetMapper and Optimized-ETDPCItemsetMapper respectively. The driver classes of optimized algorithms are similar to respective driver classes of VFPC and ETDPC except Mappers of Job2. VFPCItemsetMapper and ETDPCItemsetMapper of Job2 will be simply replaced by Optimized-VFPCItemsetMapper and Optimized-ETDPCItemsetMapper respectively.

---

**Algorithm 5** Optimized-VFPCItemsetMapper and Optimized-ETDPCItemsetMapper

| **Optimized-VFPCItemsetMapper, k ≥ 2** | **Optimized-ETDPCItemsetMapper, k ≥ 2** |
|---|---|
| Input: a block $b_i$ of database and $L_{k-1}$ | Input: a block $b_i$ of database and $L_{k-1}$ |
| key: byte offset of the line | key: byte offset of the line |
| value: a transaction $t_i \in b_i$ | value: a transaction $t_i \in b_i$ |
| 1: read $L_{k-1}$ from cache file in a prefix tree $trieL_{k-1}$ | 1: read $L_{k-1}$ from cache file in a prefix tree $trieL_{k-1}$ |
| 2: get the value of k and npass from context | 2: get the value of k and α from context |
| 3: function map(key, value) | 3: function map(key, value) |
| 4:   candidateCount = 0; | 4:   candidate threshold ct = α * $\|L_{k-1}\|$; |
| 5:   itemsetSize = k; | 5:   candidateCount = 0; |
| 6:   for (count = 1; count ≤ npass; count++) | 6:   npass = 0; |
| 7:     if(itemsetSize > k) | 7:   itemsetSize = k; |
| 8:       $trieC_k$ = non-apriori-gen($trieC_k$); | 8:   do while(candidateCount ≤ ct) |
| 9:     else | 9:     if(itemsetSize > k) |
| 10:      $trieC_k$ = apriori-gen($trieL_{k-1}$); | 10:      $trieC_k$ = non-apriori-gen($trieC_k$); |
| 11:    end if-else | 11:    else |
| 12:    $C_t$ = subset($trieC_k$, $t_i$); | 12:      $trieC_k$ = apriori-gen($trieL_{k-1}$); |
| 13:    for each candidate c ∈ $C_t$ do | 13:    end if-else |
| 14:      write (c, 1); | 14:    $C_t$ = subset($trieC_k$, $t_i$); |
| 15:    end for | 15:    for each candidate c ∈ $C_t$ do |
| 16:    candidateCount += $\|trieC_k\|$ ; | 16:      write (c, 1); |
| 17:    itemsetSize++; | 17:    end for |
| 18: end for | 18:    candidateCount += $\|trieC_k\|$ ; |
| 19: set the value of candidateCount to context | 19:    itemsetSize++, npass++; |
| 20: end function map | 20: end do-while |
| | 21: set the value of npass to context |
| | 22: end function map |

---

Both the optimized algorithms, Optimized-VFPC and Optimized-ETDPC when combines more than one passes in their multi-pass phases then the first pass invokes *apriori-gen()* to generate candidates from longest sized frequent itemsets of previous phase, and the remaining passes invoke *non-apriori-gen()* to generate next level candidates from the just immediately generated candidates of previous pass within the phase. In the both Mappers (Algorithm 5) it can be seen that *apriori-gen()* is used in the first pass and *non-apriori-gen()* in all the subsequent passes. For example, suppose a particular case in which VFPC and ETDPC and their optimized versions combine three consecutive phases *k, k+1, k+2* of SPC in a single phase. Then the Mappers of VFPC and ETDPC will generate candidates of size *k, k+1, k+2* applying *apriori-gen()* three times on frequent (k-1)-itemsets, candidate k-itemsets and candidate (k+1)-itemsets respectively while the Mappers of Optimized-VFPC and Optimized-ETDPC will first apply *apriori-gen()* to generate candidate k-itemsets from frequent (k-1)-itemsets after that *non-apriori-gen()* two times to generate candidate (k+1)-itemsets and (k+2)-itemsets from candidate k-itemsets and (k+1)-itemsets respectively. If optimized algorithms do not combine more than one passes in any phase (i.e. single pass phase) then *non-apriori-gen()* will not bring into play in that phase.

**4.3.    Analysis of Skipped-Pruning in a Multi-Pass MapReduce Phase**

Apriori is a highly computation intensive algorithm. The factors affecting its complexity are the number of transactions, number of items, average transaction width, and the user defined minimum support threshold. The computational cost of different modules (e.g. apriori-gen, subset, pruning) of sequential Apriori is described in [28] based on these factors. For Optimized-VFPC and Optimized-ETDPC algorithms, let an optimized multi-pass phase combines three consecutive passes *k, k+1, k+2*, then the dependency of computational cost of different modules in different passes can be represented as follows.

**Pass k:**

*Cost of apriori-gen operation* $\propto k*|trieL_{k-1}|*C_p$, where *k* is the size of candidate itemsets being generated, $|trieL_{k-1}|$ is the size of prefix tree containing frequent (k-1)-itemsets, and $C_p$ is the cost of pruning operation.

$C_p \propto$ *number of candidate k-itemsets generated by join step*$*(k-2)*|trieL_{k-1}|$, where *(k-2)* is the number of subset (k-1)-itemsets of k-itemsets.

*Cost of subset operation* $\propto |t_i|*|trieC_k|$, where $|t_i|$ is the width of the transaction being processed, and $|trieC_k|$ is the size of prefix tree containing candidate k-itemsets generated by apriori-gen.

**Pass k+1:**

*Cost of non-apriori-gen operation* $\propto (k+1)*|trieC_k|$, where *(k+1)* is the size of candidate itemsets being generated.

*Cost of subset operation* $\propto |t_i|*|trieC_{k+1}|$, here $|trieC_{k+1}|$ is the size of prefix tree that contains candidate (k+1)-itemsets including un-pruned (k+1)-itemsets obtained by applying non-apriori-gen.

**Pass k+2:**

*Cost of non-apriori-gen operation* $\propto (k+2)*|trieC_{k+1}|$, where (k+2) is the size of candidate itemsets being generated.

*Cost of subset operation* $\propto |t_i|*|trieC_{k+2}|$, here $|trieC_{k+2}|$ is the size of prefix tree that contains candidate (k+2)-itemsets including un-pruned (k+2)-itemsets obtained by applying non-apriori-gen.

Here it can be seen that skipped-pruning in a multi-pass phase lowers the computation cost of candidate generation but on the cost of additional un-pruned candidates. The un-pruned candidates depend on the individual datasets and *min_sup*, and cannot be quantified separately. No doubt, the un-pruned candidates increase the computation cost of self joining of candidates with itself as well as the cost of subset operation but that is not significant due to using prefix tree. The size of prefix tree will increase due to un-pruned candidates but not much since the common prefixes are stored only once in a prefix tree. Further, the *map()* method of a Mapper is invoked repeatedly for each transaction of the InputSplit assigned to that Mapper [8]. The *map()* method subsequently invokes two methods *apriori-gen()/non-apriori-gen()* and *subset()* inside it. Since the *subset()* method checks the subsets of a transaction against each candidate stored in a prefix tree, so it is fine to invoke this method for each transaction. On the other hand, the *apriori-gen()* method requires only previously generated frequent itemsets and not the transactions, so it must not be invoked for each transaction, but in MapReduce context it is invoked repeatedly. The *apriori-gen()* method further invokes a pruning method for each candidates obtained by self joining. Therefore, in a multi-pass phase, when candidates are generated from previously generated candidates, *non-apriori-gen()* is applied to skip the repeated invocation of pruning method.

The generation of candidates using *apriori-gen()* in simple multi-pass phase and using *apriori-gen()* and *non-apriori-gen()* in optimized multi-pass phase is demonstrated in Fig. 1 by an example, where we assumed a set of items $I = \{i1, i2, i3, i4, i5, i6, i7\}$ of some database $D$ and a minimum support threshold *min_sup*. As it can be seen in Fig. 1 that pass 1 and pass 2 generating frequent itemsets $L_1$ and $L_2$, are being executed in single-pass phases. Passes 3, 4 and 5 are combined in both simple and optimized multi-pass phase. For sake of the simplicity we do not mention the database and the value of *min_sup*. To generate frequent 1-itemsets, let all the items satisfy the *min_sup*. In frequent 2-itemsets generation, let itemsets *i1 i5, i2 i4* and *i4 i7* do not satisfy the *min_sup*. After generation of frequent 2-itemsets, the left branch shows candidate generation in simple multi-pass phase and the right branch shows in optimized multi-pass phase. Candidate 3-itemsets $C_3$ is the same in both types of phase since both use *apriori-gen()* for pass 3. Candidate 4 and 5-itemsets are different and distinguished as $C_4$ & $C'_4$ and $C_5$ & $C'_5$ for simple phase and optimized phase respectively. Simple phase uses *apriori-gen()* to generate $C_4$ and $C_5$ while optimized phase uses *non-apriori-gen()* to generate $C'_4$ and $C'_5$. No more candidate generation is possible further so, both stop here. Highlighted itemsets are the un-pruned candidates generated in optimized phases. It can be seen that $C_4 \subset C'_4$ and $C_5 \subset C'_5$. When both types of phases count the support for $C_3, C_4, C_5$ or $C_3, C'_4, C'_5$ and check against *min_sup*, the same set of frequent itemsets are generated at the end of phases.

### 4.4. Novelties of Proposed Algorithms

The novelties of the proposed algorithms compared to the old ones can be summarized as follows. The theoretical analysis above and the experimental results in the following section ascertain these features.

    a) VFPC and ETDPC are improvement and generalization of the efficient algorithms FPC and DPC.

    b) VFPC and ETDPC are robust and flexible while FPC and DPC are not.

    c) Optimized-VFPC and Optimized-ETDPC are further optimization over VFPC and ETDPC respectively, and outperform VFPC and ETDPC.

    d) The optimized versions are most effective for long size frequent itemset mining.

    e) All the proposed algorithms are scalable and exhibits good speedup.

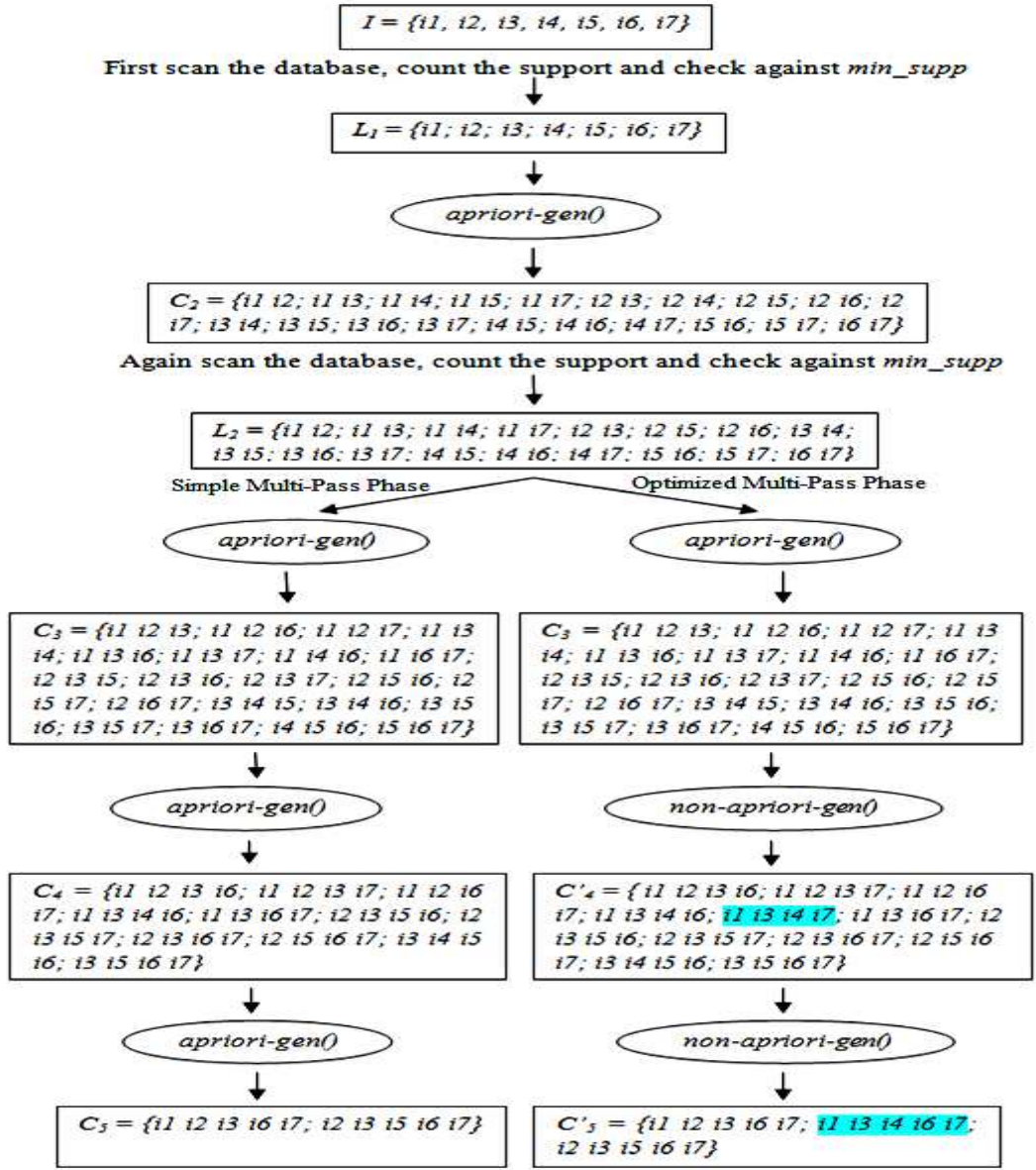

**Fig. 1.** Candidate itemsets generation in simple multi-pass phase and optimized multi-pass phase

## 5. EXPERIMENTAL EVALUATION

### 5.1. Experimental Setup and Data Sets

We have evaluated all the algorithms on local Hadoop-2.6.0 cluster installed at the Departmental Laboratory. Table 1 describes the configurations of cluster installed for experimentation. Cluster consists of 5 nodes running Ubuntu 14.04 64 bit. NameNode is exclusively configured on a virtual machine hosted on workstation while two DataNodes DN3 and DN4 are running as virtual machines hosted on another single server. DataNodes DN1 and DN2 are running on physical machines. Algorithms are implemented using Java and MapReduce 2.0 (YARN) library.

**Table 1**
Configuration of Hadoop Cluster

| Node | Node Type | # Cores | RAM | Architecture of Physical Machine |
|---|---|---|---|---|
| NameNode (NN) | Virtual | 4 | 4 GB | Intel Xenon E5-2620 @ 2.10 GHz, 12 Cores, 16 GB RAM |
| DataNode1 (DN1) | Physical | 4 | 2 GB | Intel Xenon E5504 @ 2.00 GHz, 4 Cores, 2 GB RAM |
| DataNode2 (DN2) | Physical | 4 | 2 GB | Intel Xenon E5504 @ 2.00 GHz, 4 Cores, 2 GB RAM |
| DataNode3 (DN3) | Virtual | 4 | 4 GB | Intel Xenon E5-2630 @ 2.30 GHz, 12 Cores, 32 GB RAM |
| DataNode4 (DN4) | Virtual | 4 | 4 GB | |

We have used both synthetic and real life datasets in our experiments. The synthetic dataset is c20d10k generated by IBM Generator and real datasets are chess and mushroom [29, 30]. Table 2 describes the important attributes of these datasets.

**Table 2**
Datasets used in experiments with their attributes

| Dataset | Number of Transactions ($N$) | Number of Items ($|I|$) | Average Transaction Width ($w$) |
|---|---|---|---|
| c20d10k | 10,000 | 192 | 20 |
| chess | 3196 | 75 | 37 |
| mushroom | 8124 | 119 | 23 |

**5.2. Performance Analysis**

We have evaluated the execution time of all the algorithms SPC, FPC, DPC, VFPC, ETDPC, Optimized-VFPC and Optimized-ETDPC on datasets c20d10k, chess and mushroom for varying value of minimum support. We have set $β = 60$ sec. and $α = 2.0$ for datasets c20d10k and mushroom while $α = 3.0$ for chess dataset in DPC algorithm [7] for the best possible results. InputSplit is different for different datasets. We decided it on the basis of size of dataset and the number of DataNodes available. InputSplit is configured by *setNumLinesPerSplit* method in MapReduce code. It determines the number of map tasks to be run for a job. A very small InputSplit may result into large number of mappers which may increase execution time due to parallel overhead and a large InputSplit may result into very few mappers which may not fulfill the purpose of parallel processing. All the algorithms are running with 10 and 9 map tasks on dataset c20d10k and mushroom (InputSplit is 1K lines) respectively and with 8 map tasks on chess dataset (InputSplit is 400 lines).

Figs. 2, 3 and 4 show the execution time of the various algorithms for varying value of minimum support on datasets c20d10k, chess and mushroom respectively. Figs. 2(a), 3(a) and 4(a) compare the execution of our proposed algorithms VFPC and ETDPC with the existing algorithms FPC and DPC. Figs. 2(b), 3(b) and 4(b) show the significant reduction in execution time of Optimized-VFPC and Optimized-ETDPC in comparison to VFPC and ETDPC with the decreasing value of minimum support.

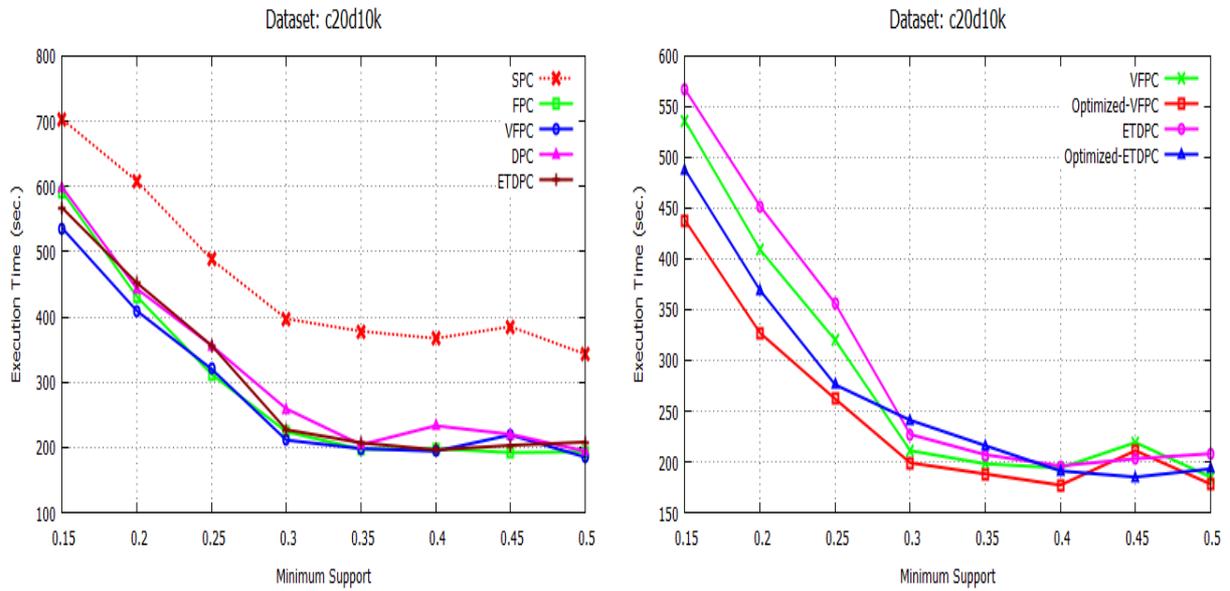

**Fig. 2.** Execution time of algorithms (a) SPC, FPC, VFPC, DPC, ETDPC (b) VFPC, Optimized-VFPC, ETDPC, Optimized-ETDPC for varying minimum support on c20d10k dataset

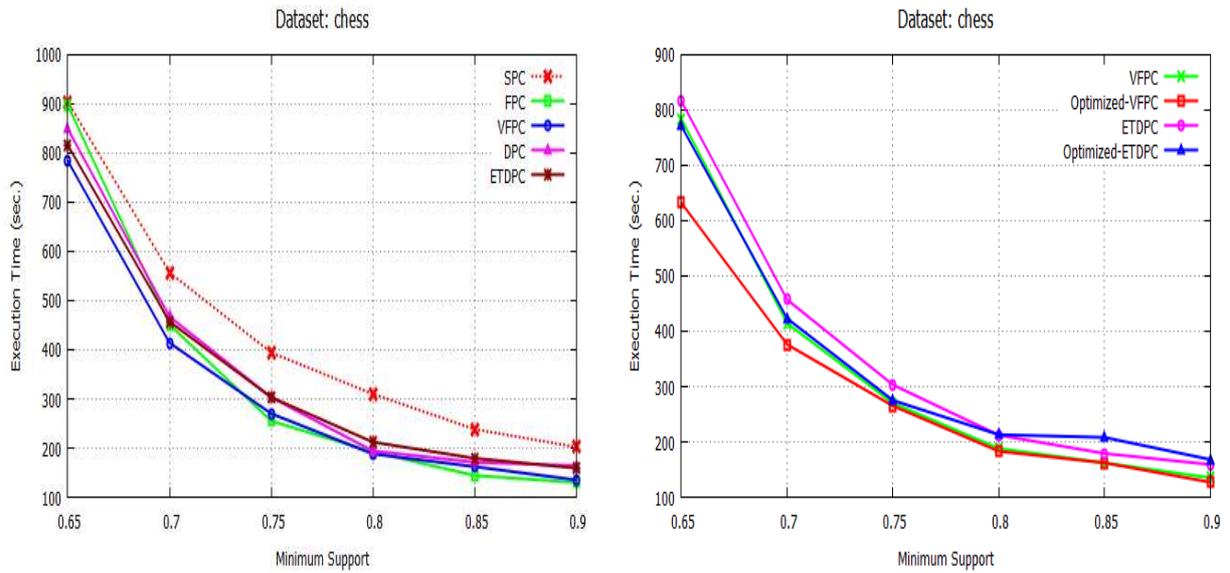

**Fig. 3.** Execution time of algorithms (a) SPC, FPC, VFPC, DPC, ETDPC (b) VFPC, Optimized-VFPC, ETDPC, Optimized-ETDPC for varying minimum support on chess dataset

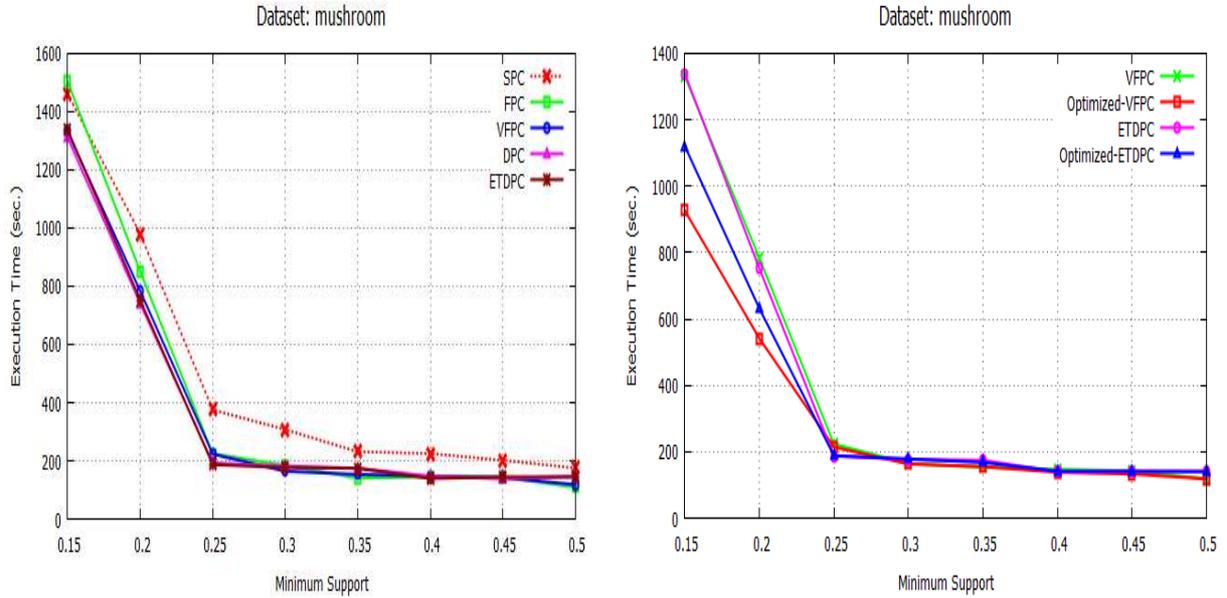

**Fig. 4.** Execution time of algorithms (a) SPC, FPC, VFPC, DPC, ETDPC (b) VFPC, Optimized-VFPC, ETDPC, Optimized-ETDPC for varying minimum support on mushroom dataset

The execution time of SPC must be an upper bound to show the efficiency of FPC, DPC, VFPC and ETDPC. All the algorithms prove this nature on the three datasets except FPC on datasets chess and mushroom (Figs. (2-4)(a)). The execution time of FPC is converging and crossing to that of SPC for the considered smallest value of minimum support on two datasets respectively (Figs. (3-4)(a)). The performance of VFPC is similar to FPC for higher value of minimum support but becomes better than FPC on lower value of minimum support. Also ETDPC performs similar to DPC with varying value of minimum support but it is more generalized algorithm than DPC. It is independent of adjusting the value of $\alpha$, as with DPC one has to adjust the value of $\alpha$ for each new dataset if it does not perform well. Optimized-VFPC and Optimized-ETDPC are showing a significant reduction in execution time in comparison to VFPC and ETDPC respectively with the decreasing value of minimum support (Figs. (2-4)(b)). Further Optimized-VFPC performs better than Optimized-ETDPC on each datasets since it supports more multi-pass phases than Optimized-ETDPC during execution as shown in next subsection.

As it has been discussed in earlier section, the computing cost is directly dependent on the length of the longest itemset being generated and the value of minimum support. Dense datasets and low value of minimum support produces longer itemsets that requires more number of passes. Further, as we decrease the value of minimum support, it increases the length of longest itemsets as well as the number of itemsets. The proposed algorithms are designed to minimize the number of passes using simple multi-pass phase and optimized multi-pass phase. At the larger value of minimum support, the required number of passes is very less and also the number of generated itemsets is very small, so all four algorithms generally execute in less than three phases. The effect of optimized multi-pass phase cannot be seen with very less number of passes. As the number of passes increases with the lower value of minimum support, the role of optimized multi-pass phase comes into the play. So, it can be seen in Figs. 2(b), 3(b) and 4(b) that when the minimum support is larger, the execution times of all four algorithms are the same. With the decreasing value of minimum support the optimized version of algorithms perform better. Following quantitative analysis critically discusses the phase-wise execution time and the number of intermediate candidate itemsets. It also adds further explanations to the performance behavior of the proposed algorithms as well as the existing algorithms.

**5.3.    Quantitative Analysis**

In order to show the validity of our proposed algorithms, we have analyzed through the quantitative perspective at the low value of minimum support. We have observed the number of generated candidates and execution time per phase on the three datasets for all the algorithms. We have represented these intermediate results of each MapReduce phase in tables to clearly visualize the effect of our applied improvements and optimizations in the algorithms. Tables 3-5 represent the breakdown of execution time of the algorithms SPC, FPC, VFPC, DPC and ETDPC on datasets c20d10k (*min_sup = 0.15*), chess (*min_sup = 0.65*) and mushroom (*min_sup = 0.15*) respectively. The value in bracket against the algorithms represents the number of phases executed in the respective algorithms.

**Table 3**
Elapsed time (sec.) of each MapReduce phase, total time (sec.) of all phases and actual execution time (sec.) on dataset c20d10k (min_sup = 0.15)

| Algorithms with number of phases | Pass 1 | Pass 2 | Pass 3 | Pass 4 | Pass 5 | Pass 6 | Pass 7 | Pass 8 | Pass 9 | Pass 10 | Pass 11 | Pass 12 | Pass 13 | Pass 14 | Total | Actual |
|---|---|---|---|---|---|---|---|---|---|---|---|---|---|---|---|---|
| SPC (14) | 16 | 18 | 24 | 32 | 48 | 70 | 91 | 83 | 51 | 34 | 22 | 18 | 16 | 21 | 544 | 703 |
| FPC (6) | 17 | 26 | 138 | | | 264 | | | 88 | | | 25 | | | 558 | 592 |
| VFPC (7) | 17 | 39 | | 80 | | 172 | | 128 | | 46 | | | 22 | | 504 | 536 |
| DPC (8) | 18 | 27 | 30 | 40 | 142 | | 180 | | 93 | | | | 23 | | 553 | 599 |
| ETDPC (8) | 18 | 27 | 30 | 84 | | 75 | 175 | | 91 | | | | 22 | | 522 | 567 |

**Table 4**
Elapsed time (sec.) of each MapReduce phase, total time (sec.) of all phases and actual execution time (sec.) on dataset chess (min_sup = 0.65)

| Algorithms with number of phases | Pass 1 | Pass 2 | Pass 3 | Pass 4 | Pass 5 | Pass 6 | Pass 7 | Pass 8 | Pass 9 | Pass 10 | Pass 11 | Pass 12 | Pass 13 | Pass 14 | Total | Actual |
|---|---|---|---|---|---|---|---|---|---|---|---|---|---|---|---|---|
| SPC (14) | 18 | 21 | 24 | 33 | 65 | 105 | 142 | 144 | 113 | 66 | 33 | 24 | 20 | 19 | 827 | 904 |
| FPC (6) | 21 | 19 | 130 | | | 476 | | | 193 | | | 27 | | | 866 | 897 |
| VFPC (7) | 19 | 25 | | 89 | | 255 | | 254 | | 84 | | | 20 | | 746 | 784 |
| DPC (9) | 21 | 21 | 27 | 36 | 171 | | 151 | | 262 | | 90 | | 21 | | 800 | 849 |
| ETDPC (8) | 20 | 20 | 25 | 35 | 167 | | 149 | | 338 | | | | 21 | | 775 | 816 |

**Table 5**
Elapsed time (sec.) of each MapReduce phase, total time (sec.) of all phases and actual execution time (sec.) on dataset mushroom (min_sup = 0.15)

| Algorithms with number of phases | Pass 1 | Pass 2 | Pass 3 | Pass 4 | Pass 5 | Pass 6 | Pass 7 | Pass 8 | Pass 9 | Pass 10 | Pass 11 | Pass 12 | Pass 13 | Pass 14 | Pass 15 | Pass 16 | Total | Actual |
|---|---|---|---|---|---|---|---|---|---|---|---|---|---|---|---|---|---|---|
| SPC (16) | 23 | 24 | 30 | 56 | 94 | 150 | 207 | 225 | 200 | 143 | 86 | 45 | 30 | 24 | 22 | 21 | 1380 | 1460 |
| FPC (7) | 22 | 24 | 377 | | | 580 | | | 382 | | | 53 | | | 25 | | 1463 | 1503 |
| VFPC (7) | 21 | 52 | | 151 | | 341 | | 434 | | | 266 | | | | 26 | | 1291 | 1330 |
| DPC (9) | 23 | 23 | 28 | 53 | 222 | | 210 | 430 | | | 246 | | | | 22 | | 1257 | 1311 |
| ETDPC (10) | 21 | 25 | 30 | 56 | 96 | 152 | 210 | 432 | | | 243 | | | | 23 | | 1288 | 1338 |

In Tables 3-5, it can be seen that FPC and VFPC executed with nearly the same number of phases (either 6 or 7) and similar occurred with DPC and ETDPC. With respect to execution time, ETDPC performs almost similar to DPC whereas VFPC performs better than FPC and never converges to SPC. As it can be seen in Table 5 that FPC suffers with overloaded candidates in the phase that combines passes 3 to 5 and takes longer time in that phase. Consequently, in spite of only seven phases taken by FPC compared to sixteen phases taken by SPC, the execution of FPC is greater than that of SPC. Further, FPC and VFPC both executed in seven phases but have a running time difference of more than 200 seconds (Table 5). For some dataset, it may be possible that FPC is stuck with such overloaded phase and terminates in extremely large time. Also in Table 4, FPC completes in six phases while VFPC in seven phases, in spite of that FPC takes more time than VFPC. All these happen only due the ad-hoc combination of passes in a phase by FPC. In the tables representing execution time, total time is the sum of elapsed time of all phases whereas actual time is final execution time obtained at the completion of the algorithms. More the number of phases an algorithm takes, more the gap increases between total time and actual time. The proposed algorithms are better because they complete execution in less number of phases and these phases combine passes in an intelligent way. Before investigating the number of candidates and elapsed time per phase of the proposed algorithms, we have summarized the number of frequent itemsets generated in each passes of Apriori on the three datasets. Table 6 lists the number of frequent itemsets in each passes of Apriori generated from the datasets c20d10k, chess and mushroom at minimum support 0.15, 0.65 and 0.15 respectively. It can be seen that the number of frequent itemsets in latter passes are very less than the earlier and middle passes.

**Table 6**
Number of frequent k-itemsets $L_k$ (k ≥ 1) generated from datasets c20d10k (min_sup = 0.15), chess (min_sup = 0.65) and mushroom (min_sup = 0.15)

| Datasets | $L_1$ | $L_2$ | $L_3$ | $L_4$ | $L_5$ | $L_6$ | $L_7$ | $L_8$ | $L_9$ | $L_{10}$ | $L_{11}$ | $L_{12}$ | $L_{13}$ | $L_{14}$ | $L_{15}$ |
|---|---|---|---|---|---|---|---|---|---|---|---|---|---|---|---|
| c20d10k | 38 | 319 | 1349 | 3545 | 6352 | 8163 | 7615 | 5230 | 2607 | 918 | 217 | 31 | 3 | 0 | 0 |
| chess | 29 | 307 | 1716 | 5992 | 13927 | 22442 | 25713 | 21111 | 12329 | 5027 | 1384 | 240 | 19 | 0 | 0 |
| mushroom | 48 | 530 | 2510 | 6751 | 12372 | 17008 | 18745 | 16887 | 12290 | 7052 | 3094 | 1001 | 224 | 31 | 2 |

Tables 7-9 show the number of candidates generated in each MapReduce phase of the algorithms SPC, VFPC, Optimized-VFPC, ETDPC and Optimized-ETDPC on the three datasets. We have omitted phase-1 here since it does not generate any candidates. Tables 10-12 show the elapsed time of each MapReduce phase of the algorithms VFPC, Optimized-VFPC, ETDPC and Optimized-ETDPC on the three datasets. If we observe the number of candidates and the elapsed time of each phase in corresponding tables (Tables 7-9 and Tables 10-12) on respective datasets, a significant reduction in elapsed time can be seen in phases of Optimized-VFPC and Optimized-ETDPC in spite of increased number of candidates. So it can be seen that the optimized multi-pass phase executes faster than simple multi-pass phase. In Tables 7-9, there is more number of candidates generated in the multi-pass phases of Optimized-VFPC in comparison to VFPC. Similar happens between Optimized-ETDPC and ETDPC, if they combine the similar passes in a multi-pass phase. The increased number of candidates is due to the un-pruned candidates which are generated due to skipping pruning step. In Tables 10-12, the elapsed time of multi-pass phases in Optimized-VFPC is significantly less than that of VFPC in spite of the increased number of candidates in multi-pass phases of Optimized-VFPC. Similar happens between Optimized-ETDPC and ETDPC, if they combine the similar passes in a multi-pass phase. The reduction in elapse time is due to the skipped-pruning in multi-pass phases. Further, it can be seen in Tables 10-12 that Optimized-VFPC performs better than Optimized-ETDPC since it supports more number of optimized multi-pass phases than that of Optimized-ETDPC and also the total number of phases executed in Optimized-VFPC is less than that of Optimized-ETDPC.

**Table 7**
Number of candidates generated in each MapReduce phase on dataset c20d10k (min_sup = 0.15)

| Algorithms | Pass 2 | Pass 3 | Pass 4 | Pass 5 | Pass 6 | Pass 7 | Pass 8 | Pass 9 | Pass 10 | Pass 11 | Pass 12 | Pass 13 |
|---|---|---|---|---|---|---|---|---|---|---|---|---|
| SPC | 703 | 2602 | 3651 | 6391 | 8170 | 7616 | 5230 | 2607 | 918 | 217 | 31 | 2 |
| VFPC | 9139 | | 10449 | | 15769 | | 7873 | | 1168 | | | |
| Optimized-VFPC | 9139 | | 12689 | | 17876 | | 8459 | | 1245 | | | |
| ETDPC | 703 | 2602 | 10449 | | 8140 | 12850 | | 3775 | | | | |
| Optimized-ETDPC | 703 | 2602 | 12689 | | 8140 | 14161 | | 4189 | | | | |

**Table 8**
Number of candidates generated in each MapReduce phase on dataset chess (min_sup = 0.65)

| Algorithms | Pass 2 | Pass 3 | Pass 4 | Pass 5 | Pass 6 | Pass 7 | Pass 8 | Pass 9 | Pass 10 | Pass 11 | Pass 12 | Pass 13 |
|---|---|---|---|---|---|---|---|---|---|---|---|---|
| SPC | 406 | 2179 | 6777 | 15231 | 23728 | 26586 | 21537 | 12469 | 5051 | 1387 | 243 | 22 |
| VFPC | 4060 | | 25415 | | 53780 | | 34866 | | 6727 | | | |
| Optimized-VFPC | 4060 | | 28275 | | 58454 | | 38117 | | 7634 | | | |
| ETDPC | 406 | 2179 | 6777 | 42899 | | 26586 | 43417 | | | | | |
| Optimized-ETDPC | 406 | 2179 | 6777 | 15231 | 23728 | 26586 | 38117 | | 7634 | | | |

**Table 9**
Number of candidates generated in each MapReduce phase on dataset mushroom (min_sup = 0.15)

| Algorithms | Pass 2 | Pass 3 | Pass 4 | Pass 5 | Pass 6 | Pass 7 | Pass 8 | Pass 9 | Pass 10 | Pass 11 | Pass 12 | Pass 13 | Pass 14 | Pass 15 |
|---|---|---|---|---|---|---|---|---|---|---|---|---|---|---|
| SPC | 1128 | 4599 | 7774 | 12586 | 17095 | 18771 | 16888 | 12290 | 7052 | 3097 | 1001 | 224 | 31 | 2 |
| VFPC | 18424 | | 22388 | | 36016 | | 29178 | | 11405 | | | | | 2 |
| Optimized-VFPC | 18424 | | 29676 | | 38373 | | 29462 | | 11409 | | | | | 2 |
| ETDPC | 1128 | 4599 | 7774 | 12586 | 17095 | 18771 | 29178 | | 11407 | | | | | |
| Optimized-ETDPC | 1128 | 4599 | 7774 | 12586 | 17095 | 18771 | 29462 | | 11411 | | | | | |

**Table 10**
Elapsed time (sec.) of each MapReduce phase, total time (sec.) of all phases and actual execution time (sec.) on dataset c20d10k (min_sup = 0.15)

| Algorithms with number of phases | Pass 1 | Pass 2 | Pass 3 | Pass 4 | Pass 5 | Pass 6 | Pass 7 | Pass 8 | Pass 9 | Pass 10 | Pass 11 | Pass 12 | Pass 13 | Pass 14 | Total | Actual |
|---|---|---|---|---|---|---|---|---|---|---|---|---|---|---|---|---|
| VFPC (7) | 17 | 39 | | 80 | | 172 | | 128 | | 46 | | | | 22 | 504 | 536 |
| Optimized-VFPC (7) | 17 | 35 | | 63 | | 117 | | 98 | | 45 | | | | 23 | 398 | 438 |
| ETDPC (8) | 18 | 27 | 30 | 84 | | 75 | | 175 | | 91 | | | | 24 | 524 | 567 |
| Optimized-ETDPC (8) | 18 | 27 | 30 | 65 | | 80 | | 126 | | 75 | | | | 23 | 444 | 488 |

**Table 11**
Elapsed time (sec.) of each MapReduce phase, total time (sec.) of all phases and actual execution time (sec.) on dataset chess (min_sup = 0.65)

| Algorithms with number of phases | Pass 1 | Pass 2 | Pass 3 | Pass 4 | Pass 5 | Pass 6 | Pass 7 | Pass 8 | Pass 9 | Pass 10 | Pass 11 | Pass 12 | Pass 13 | Pass 14 | Total | Actual |
|---|---|---|---|---|---|---|---|---|---|---|---|---|---|---|---|---|
| VFPC (7) | 19 | 25 | | 89 | | 255 | | 254 | | 84 | | | | 20 | 746 | 784 |
| Optimized-VFPC (7) | 20 | 23 | | 75 | | 189 | | 192 | | 76 | | | | 22 | 597 | 632 |
| ETDPC (8) | 20 | 20 | 25 | 35 | 167 | | 149 | | 338 | | | | | 21 | 775 | 816 |
| Optimized-ETDPC (10) | 19 | 21 | 26 | 39 | 63 | 108 | 149 | 194 | | 76 | | | | 22 | 717 | 771 |

**Table 12**
Elapsed time (sec.) of each MapReduce phase, total time (sec.) of all phases and actual execution time (sec.) on dataset mushroom (min_sup = 0.15)

| Algorithms with number of phases | Pass 1 | Pass 2 | Pass 3 | Pass 4 | Pass 5 | Pass 6 | Pass 7 | Pass 8 | Pass 9 | Pass 10 | Pass 11 | Pass 12 | Pass 13 | Pass 14 | Pass 15 | Pass 16 | Total | Actual |
|---|---|---|---|---|---|---|---|---|---|---|---|---|---|---|---|---|---|---|
| VFPC (7) | 21 | 52 | | 151 | | 341 | | 434 | | 266 | | | | | 26 | | 1291 | 1330 |
| Optimized-VFPC (7) | 19 | 52 | | 108 | | 231 | | 261 | | 187 | | | | | 25 | | 883 | 926 |
| ETDPC (10) | 21 | 25 | 30 | 56 | 96 | 152 | 210 | 432 | | 243 | | | | | 23 | | 1288 | 1338 |
| Optimized-ETDPC (10) | 21 | 26 | 30 | 58 | 98 | 154 | 205 | 268 | | 186 | | | | | 22 | | 1068 | 1120 |

### 5.4. Scalability and Speedup

Scalability and Speedup test were carried out for our proposed algorithms on dataset c20d10k for a fixed value of minimum support. For scalability test we executed the four algorithms with 10 Mappers on increasing size of dataset c20d10 (*min_sup = 0.25*). Fig. 5(a) shows the scalability of VFPC, Optimized-VFPC, ETDPC and Optimized-ETDPC. We scaled up the InputSplit with increasing size of dataset so that the number of total map tasks remains constant. Speedup is calculated for the four algorithms on increasing number of DataNodes. Speedup is the ratio of sequential execution time (SET) and parallel execution time (PET). Sequential execution time is obtained on single DataNode and parallel execution time is obtained by adding more DataNodes. Speedup achieved by a parallel algorithm depends on the serial portion and parallel portion of the program. A parallel algorithm having more parallel portion over sequential portion achieves more speedup. We executed the algorithms with 10 Mappers for different number of DataNodes on dataset c20d200k (i.e. c20d10k with 200K lines) with *min_sup = 0.40*. Fig. 5(b) shows the speedup of VFPC, Optimized-VFPC, ETDPC and Optimized-ETDPC for increasing number of DataNodes.

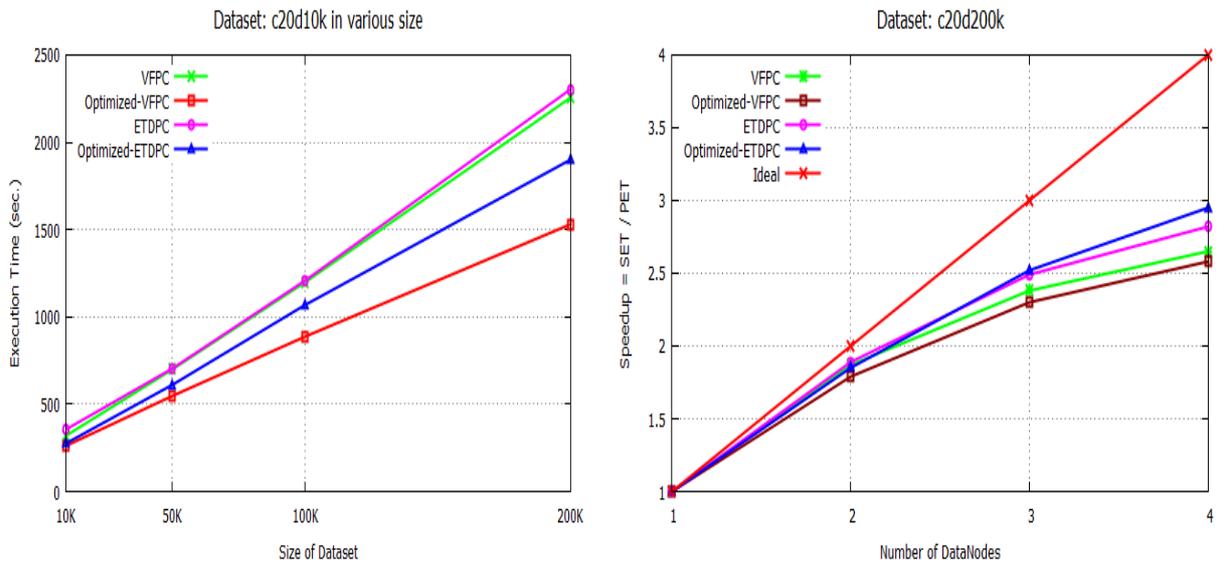

**Fig. 5.** (a) Execution time (sec.) on increasing size of dataset (b) Speedup on increasing number of DataNodes

### 6. CONCLUSIONS

We have addressed the problem of performance optimization of MapReduce based Apriori algorithm by reducing the number of passes, and proposed four algorithms VFPC, ETDPC, Optimized-VFPC and Optimized-ETDPC based on the combined passes of Apriori in a single MapReduce phase. VFPC and ETDPC are reliant on the number of candidates in two preceding consecutive phases and the elapsed time of these phases respectively. They are more robust and flexible than existing algorithms FPC and DPC against new datasets as well as on clusters with different computing capacity. Optimized-VFPC and Optimized-ETDPC optimize the multi-pass phases of VFPC and ETDPC respectively by applying a combination *non-apriori-gen()* and *apriori-gen()* methods. We have introduced a method *non-apriori-gen()* which generates candidates by skipping pruning step. In an optimized multi-pass phase, the traditional *apriori-gen()* is applied in the first pass and the *non-apriori-gen()* in the remaining passes. The skipped-pruning produces additional un-pruned candidates that slightly increase the size of prefix tree used to store the candidate itemsets. Experimental results reveal that counting cost of un-pruned candidates is less significant than the reduction in computation due to skipped-pruning. Experiments are carried out on the both synthetic and real life datasets on the varying minimum support. It has been found that VFPC outperforms FPC and also do not suffer with overloaded phase. With DPC one has to adjust the parameters for new datasets but ETDPC is free from such adjustment and with the performance similar to DPC. Optimized-VFPC and Optimized-ETDPC outperforms both VFPC and ETDPC on each dataset. Quantitative analysis on individual phases of all algorithms reveals that in spite of increased number of candidates, elapsed time of optimized multi-pass phases is significantly less than that of simple multi-pass phases. Scalability and speedup test shows that all the four algorithms are scalable with increasing size of dataset and achieve good speedup with increasing number of nodes.

**ACKNOWLEDGEMENT**

The authors would like to thank the anonymous reviewers for giving valuable comments and suggestions to revise the manuscript in the present form.

**Sudhakar Singh** received his M.C.A. and Ph.D. degree from Department of Computer Science, Institute of Science, Banaras Hindu University, India in 2010 and 2016 respectively. He is working as an Assistant Professor in the same University since 2016. His research interests include Scalable Algorithms for Big Data, and Distributed Computing.

**Rakhi Garg** is an Associate Professor at Mahila Maha Vidyalaya, Banaras Hindu University, India. She received her M.Sc. in 1997 and Ph. D. in 2012 from Department of Computer Science, Institute of Science, Banaras Hindu University. Her research interests include Data Mining, Web Mining, and Cluster Computing.

**P. K. Mishra** is Professor at Department of Computer Science, Institute of Science, Banaras Hindu University, India. He is also a Principal Investigator of the research projects at DST Centre for Interdisciplinary Mathematical Sciences, Banaras Hindu University. He is a senior member of IEEE. His research interests include Computational Complexity, Data Mining, High Performance Computing and VLSI Algorithms.